\documentclass[a4paper]{jpconf}
\usepackage{graphicx}
\usepackage{footmisc}
\usepackage{amssymb}
\usepackage{siunitx}
\def\beq{\begin{equation}}
\def\eeq{\end{equation}}

\def\beqn{\begin{eqnarray}} 
\def\eeqn{\end{eqnarray}}

\def\bea{\begin{eqnarray}}
\def\eea{\end{eqnarray}}

\renewcommand{\thefootnote}{\fnsymbol{footnote}}

\begin{document}
\title{Global extraction of the parton-to-pion fragmentation functions at NLO accuracy in QCD}

\author{R. J. Hern\'andez-Pinto$^1$\footnote[1]{\hspace{1mm}Speaker}, M. Epele$^2$, D. de Florian$^{3}$, R. Sassot$^{4}$ and M. Stratmann$^5$}
\address{$^1$ Facultad de Ciencias F\'isico-Matem\'aticas, Universidad Aut\'onoma de Sinaloa, Ciudad Universitaria, CP 80000, Culiac\'an, Sinaloa, M\'exico}
\address{$^2$ Instituto de F\'isica La Plata, UNLP, CONICET Departamento de F\'isica, Facultad de Ciencias Exactas, Universidad de La Plata, C.C. 69, La Plata, Argentina}
\address{$^3$ International Center for Advanced Studies (ICAS), UNSAM, Campus Miguelete, 25 de Mayo y Francia, (1650) Buenos Aires, Argentina}
\address{$^4$ Departamento de F\'isica and IFIBA, Facultad de Ciencias Exactas y Naturales, Universidad de Buenos Aires, Ciudad Universitaria, Pabell\'on 1 (1428) Buenos Aires, Argentina}
\address{$^5$ Institute for Theoretical Physics, University of T\"ubingen, Auf der Morgenstelle 14, 72076 T\"ubingen, Germany}

\ead{roger@uas.edu.mx}

\begin{abstract}

In this review, we discuss the results on the parton-to-pion fragmentation functions obtained in a
combined NLO fit to data of single-inclusive hadron production in electron-positron annihilation,
proton-proton collisions, and lepton-nucleon deep-inelastic scattering. A more complete discussion can be found in Ref.~\cite{ref:dss01}

\end{abstract}

\section{Introduction}

Fragmentation functions (FFs) are fundamental objects which describe the collinear transition of a
quark $i$ into a hadron $H$ with a fraction $z$ of its momentum, and it is usually named as $D^H_i(z)$.
These FFs can only be obtained by performing global fits. However, since the 
FFs are non perturbative objects, they cannot be computed from first principles and 
they need to be extracted by fitting the experimental data of different kind of processes. However, the 
scale dependence of the FFs can be obtained in pertubative QCD (pQCD) and can 
be determined by renormalization group equations, similar to those for parton densities (PDF).

In here we present the results of Ref.~\cite{ref:dss01} where we obtained an updated set of parton-to-pion FFs
with the determination of their uncertanties by applying the IH method in light of all the newly 
available, precise experimental results in SIA, SIDIS, and $pp$ collisions.
This will allow us to scrutinize the consistency of the information on FFs extracted across the different
hard scattering processes, i.e., to validate the fundamental notion of universality, 
which is at the heart of any pQCD calculation based on the factorization of short- 
and long-distance physics \cite{ref:fact}.

\section{Functional Form and Fit Parameters \label{sec:funcform}}

We parametrize the hadronization of a parton of flavor $i$ 
into a positively charged pion at an initial scale of $Q_0=1\,\mathrm{GeV}$ as
\begin{equation}
\label{eq:ff-input}
D_i^{\pi^+}\!(z,Q_0) =
\frac{N_i z^{\alpha_i}(1-z)^{\beta_i} [1+\gamma_i (1-z)^{\delta_i}] }
{B[2+\alpha_i,\beta_i+1]+\gamma_i B[2+\alpha_i,\beta_i+\delta_i+1]}\;.
\end{equation}
Here, $B[a,b]$ denotes the Euler Beta-function, and the 
$N_i$ in (\ref{eq:ff-input}) are chosen in such a way that
they represent the contribution of $z D_i^{\pi^+}$ to the momentum
sum rule.

The improved experimental information now allows us to impose less 
constraints on the parameter space. More specifically, as before we still have to 
assume isospin symmetry, i.e.,
\begin{equation}
\label{eq:su2constraint}
D_{\bar{u}}^{\pi^+}=D_{d}^{\pi^+}\;,
\end{equation}
and we need to relate the total $u$-quark and $d$-quark FFs 
by a global, $z$-independent factor $N_{d+\bar{d}}$,
\begin{equation}
\label{eq:su2breaking}
D_{d+\bar{d}}^{\pi^+} = N_{d+\bar{d}} D_{u+\bar{u}}^{\pi^+}\;,
\end{equation}
which quantifies any charge symmetry violation found in the fit.
The fragmentation of a strange quark into a pion
is now related to the unfavored FFs 
in Eq.~(\ref{eq:su2constraint}) by 
\begin{equation}
\label{eq:su3breaking}
D_s^{\pi^+}=D_{\bar{s}}^{\pi^+} = N_s z^{\alpha_s} D_{\bar{u}}^{\pi^+} \;,
\end{equation}
rather than just using a constant as in the DSS analysis. Besides, the charm- and bottom-to-pion 
FFs no longer assume $\gamma_c=\gamma_b=0$ in Eq.~(\ref{eq:ff-input}) but can now exploit the 
full flexibility of the ansatz. As in the DSS and all 
other analyses \cite{ref:dss,ref:dss2,ref:eta,ref:kretzer,ref:other-ffs}, we include heavy flavor FFs 
discontinuously as massless partons in the QCD scale evolution 
above their $\overline{\rm{MS}}$ ``thresholds", $Q=m_{c,b}$,
with $m_c$ and $m_b$ denoting the mass of the charm and bottom quark,
respectively. 

In total we now have 28 free fit parameters describing our updated FFs for quarks,
antiquarks, and gluons into positively charged pions.
The corresponding FFs for negatively charged pions are obtained by charge conjugation 
and those for neutral pions by assuming $D_i^{\pi^0}= [D_i^{\pi^+}+D_i^{\pi^-}]/2$.

\section{Data Selection \label{sec:datasets}}

For the global fit, we use of all the available experimental information on single-inclusive
charged and neutral pion production in SIA, SIDIS, and hadron-hadron collisions.

Compared to the data sets already used in the DSS global analysis \cite{ref:dss}, we include the new
results from {\sc BaBar} \cite{ref:babardata} and {\sc Belle} \cite{ref:belledata} in SIA 
at a c.m.s.\ energy of $\sqrt{S}\simeq 10.5\,\mathrm{GeV}$.
Both sets are very precise, with relative uncertainties of about $2-3\%$, and reach
all the way up to pion momentum fractions $z$ close to one, well beyond of what has been measured so far. 
As customary, we
limit ourselves to data with $z\ge 0.1$ to avoid any potential impact from kinematical
regions where finite, but neglected, hadron mass corrections,
proportional to $M_{\pi}/(S z^2)$, might become of any importance \cite{ref:dss,ref:kretzer,ref:other-ffs}. 

In case of SIDIS, we replace the preliminary multiplicity data from {\sc Hermes} \cite{ref:hermes-old}
by their released final results \cite{ref:hermesmult}.
More specifically, we use the data for charged pion multiplicities as a function of momentum transfer $Q^2$ 
in four bins of $z$ taken on both a proton and a deuteron target. 
In addition, we include the still preliminary multiplicity data for $\pi^{\pm}$ 
from the {\sc Compass} Collaboration \cite{ref:compassmult}, which are given as a function of $z$ in bins 
of $Q^2$ and the initial-state momentum fraction $x$. 
In addition, for the SIDIS data sets we do not have to impose any cuts on both data sets to accommodate 
them in the global analysis.

Finally, we add a couple of new sets of data for inclusive high-$p_T$ pion production in $pp$ collisions 
to the results from the {\sc Phenix} experiment \cite{ref:phenixdata} already included in the DSS analysis. 
Most noteworthy are the first results for neutral pions from the {\sc Alice} Collaboration 
at CERN-LHC \cite{ref:alicedata}, covering 
unprecedented c.m.s.\ energies of up to $7\,\mathrm{TeV}$. In addition, we
add {\sc Star} data taken at $\sqrt{S}=200\,\mathrm{GeV}$ in various rapidity intervals
for both neutral and charged pion production and for the $\pi^-/\pi^+$ ratio
\cite{ref:starcharged06,ref:stardata09,ref:starratio11,ref:stardata13}. 
It is worth mentioning that
it turns out that a good global fit of RHIC and LHC $pp$ data, along with all the other world data, 
can only be achieved if one imposes a cut on the 
minimum $p_T$ of the produced pion of about $5\,\mathrm{GeV}$. 

\section{Fit Procedure and Uncertainty Estimates \label{sec:uncertainties}}

The 28 free parameters describing the updated parton-to-pion FFs 
in Eq.~(\ref{eq:ff-input}) at the chosen input scale of $1\,\mathrm{GeV}$ are again
determined from a standard $\chi^2$ minimization where
\begin{equation}
\label{eq:chi2}
\chi^2=\sum_{i=1}^N  \left[
\left( \frac{1-{\cal{N}}_i}{\delta{\cal{N}}_i}  \right)^2 +
\sum_{j=1}^{N_i}
\frac{({\cal{N}}_i T_j-E_j)^2}{\delta E_j^2} \right],
\end{equation}
for $i=1,\ldots,N$ data sets, each contributing with $N_i$ data points.
$E_j$ is the measured value of a given observable,
$\delta E_j$ the error associated with this measurement, and
$T_j$ is the corresponding theoretical estimate for a
given set of parameters in Eq.~(\ref{eq:ff-input}).
In this new fit,
we derive the optimum normalization shifts
analytically from the condition $\partial \chi^2/\partial {\cal{N}}_i=0$,
which yields
\begin{equation}
\label{eq:normshift}
{\cal{N}}_i = \frac{ \sum_{j=1}^{N_i} \frac{\delta{\cal{N}}_i^2}{\delta E_j^2} T_j E_j +1}
{1+ \sum_{j=1}^{N_i}  \frac{\delta{\cal{N}}_i^2}{\delta E_j^2} T_j^2  }\;.
\end{equation}
Here, $\delta {\cal{N}}_i$ denotes the quoted experimental normalization uncertainty
for data set $i$.

Now, in order to estimates the uncertainties we use the IH method~\cite{ref:ih}. The main idea of the method 
is to assume a quadratic behavior of the $\chi^2$ hyper-surface
of parameter displacements and to express the $\chi^2$ increment from its minimum value 
in terms of combinations of fit parameters that maximize the variation. 
These sets
correspond to fixed displacements along the eigenvector directions of the Hessian matrix. 
To define the eigenvector sets one has to choose a tolerance parameter 
$\Delta \chi^2$ for the increment in $\chi^2$ which is still
acceptable in the global fit.
Here we proceed as follows: the tolerances for the eigenvector sets
corresponding to $68\%$ and $90\%$ confidence level (C.L.) intervals
are determined from the Gaussian probability density function for a
$\chi^2$ distribution with $k$ degrees of freedom (d.o.f.):
\begin{equation}
P_k(x) = \frac{x^{k/2-1} e^{-x/2}}{\Gamma(k/2)2^{k/2}}\;.
\end{equation}
The $\Delta \chi^2$ related to the $68^{th}$ and $90^{th}$ percentiles are 
then obtained by solving
$\int_0^{\chi^2+\Delta \chi^2} d\chi^2 P_k(\chi^2) = 0.68$ and $0.90$, respectively.

Finally, we choose the NLO set of PDFs from the MSTW group \cite{ref:mstw} and the
corresponding uncertainty estimates in computations of the SIDIS and $pp$ cross sections.
For consistency, we also fix the strong coupling $\alpha_{\rm S}$ to the values obtained in 
the MSTW fit. 

\section{Results}

In this section we present and discuss the results of our global
analysis of parton-to-pion fragmentation functions. 
First, we present the obtained fit parameters,
normalization shifts, and individual $\chi^2$ values. 
Next, the obtained $D_i^{\pi^{+}}(z,Q^2)$ and their uncertainties
are shown and compared to the results of the DSS fit.

\subsection{Parton-To-Pion Fragmentation Functions \label{sec:ff-results}}

Table~\ref{tab:nlopionpara} reveals already a notable difference to one of the findings of the DSS analysis
which preferred an unexpectedly sizable breaking of the charge symmetry between
$u+\bar{u}$ and $d+\bar{d}$ FFs of about $10\%$ \cite{ref:dss}, within large uncertainties though. 
Now, with much improved experimental information on charged pion multiplicities
both from {\sc Hermes} \cite{ref:hermesmult} and {\sc Compass} \cite{ref:compassmult} 
and new data on the ratio $\pi^-/\pi^+$ in $pp$ collisions from {\sc Star} \cite{ref:starratio11}, 
the parameter $N_{d+\bar{d}}$ in Eq.~(\ref{eq:su2breaking}) prefers to stay
very close to unity, i.e., very little or no breaking. 
\begin{table}[th!]
\begin{center}
\begin{tabular}{|cccccc|}
\hline
\hline
flavor $i$ &$N_i$ & $\alpha_i$ & $\beta_i$ &$\gamma_i$ &$\delta_i$\\
\hline
$u+\overline{u}$ & 0.387&-0.388& 0.910& 7.15& 3.96\\
$d+\overline{d}$ & 0.388&-0.388& 0.910& 7.15& 3.96\\
$\overline{u}=d$ & 0.105& 1.649& 3.286&49.95& 8.67\\
$s+\overline{s}$ & 0.273& 1.449& 3.286&49.95& 8.67\\
$c+\overline{c}$ & 0.306& 1.345& 5.519&19.78&10.22\\
$b+\overline{b}$ & 0.372&-0.127& 4.490&24.49&12.80\\
$g$                     & 0.260& 2.552& 6.194&87.06&20.36\\
\hline
\hline
\end{tabular}
\end{center}
\caption{\label{tab:nlopionpara}Parameters describing the NLO FFs for positively charged
pions, $D_i^{\pi^+}(z,Q_0)$,
in Eq.~(\ref{eq:ff-input}) in the $\overline{\mathrm{MS}}$ scheme at the input scale $Q_0=1\,\mathrm{GeV}$.
Results for the charm and bottom FFs refer to
$Q_0=m_c=1.43\,\mathrm{GeV}$ and
$Q_0=m_b=4.3\,\mathrm{GeV}$, respectively.}
\end{table}

\begin{figure*}[hbt!]
\vspace*{-0.4cm}
\begin{center}
\includegraphics[scale=0.37]{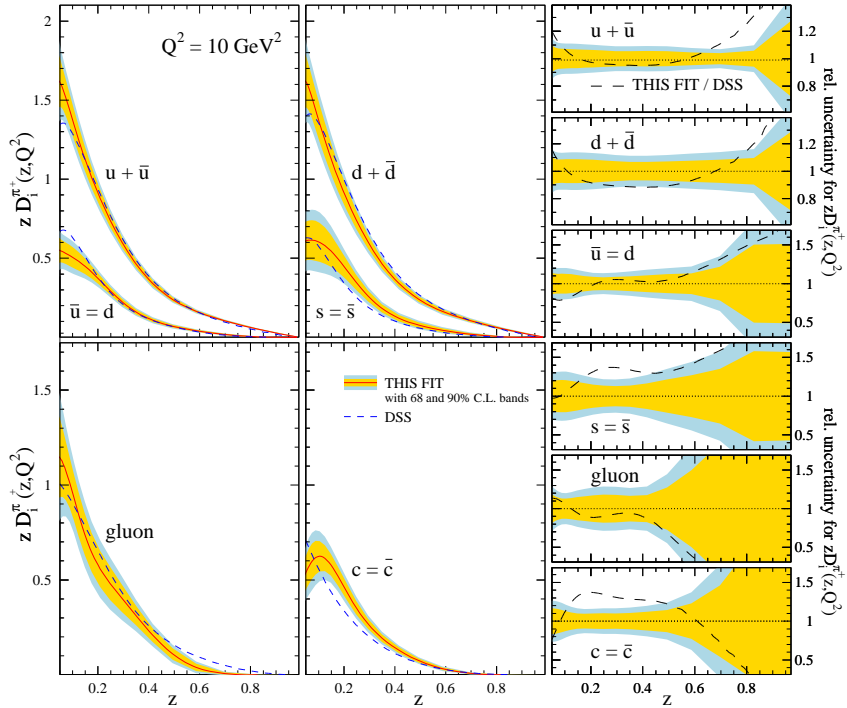}
\end{center}
\vspace*{-0.5cm}
\caption{The individual FFs for positively charged pions $zD_i^{\pi^{+}}(z,Q^2)$ at
$Q^2=10\,\mathrm{GeV}^2$ along with uncertainty estimates at $68\%$ and $90\%$ C.L.\
indicated by the inner and outer shaded bands, respectively.
The panels on the right-hand-side show the corresponding relative uncertainties.
Also shown is a comparison to the previous global analysis by DSS \cite{ref:dss} (dashed lines).
\label{fig:ff-at-10}}
\end{figure*}

In Fig.~\ref{fig:ff-at-10} we present the new parton-to-pion FFs at $Q^2 = 10$ GeV$^2$. 
As can be inferred, for the light quark flavors the old DSS results are either close to
the updated fit or within its $90\%$ C.L.\ uncertainty band.
The best determined pion FFs is $D_{u+\bar{u}}^{\pi^{+}}$, where the relative uncertainties
are below $10\%$ at $90\%$ C.L.\ throughout most of the relevant $z$ range. 
Only for $z\gtrsim 0.8$ the errors rapidly
increase because of the lack of experimental constraints in this region.
The corresponding uncertainties for $D_{d+\bar{d}}^{\pi^{+}}$ turn out to be slightly larger
as they also include possible violations of SU(2) charge symmetry through Eq.~(\ref{eq:su2breaking}).
Bigger deviations from the DSS analysis are found for both the gluon and the charm FFs.
In the latter case, this is driven by the greater flexibility of the functional form,
five fit parameters rather than three. 
The significantly reduced $D_g^{\pi^{+}}$ as compared to the DSS fit is a result of the
new {\sc Alice} $pp$ data \cite{ref:alicedata}, which have a strong preference for less pions from
gluon fragmentation for basically all values of $z$. Similar conclusions are obtained for $Q^2= M_{\it Z}^2$.

\begin{table}[bth!]
\begin{center}
\begin{tabular}{|lcccc|}
\hline 
\hline
experiment&  &  & \# data in fit & $\chi^2$ 
         \\\hline
{\sc Tpc} \cite{ref:tpcdata}  &             &   & 44 & 37.1 \\
{\sc Tasso} \cite{ref:tassodata}  &      &      & 18 &  54.7    \\
{\sc Sld} \cite{ref:slddata}  &   &   & 79 & 62.1 \\
{\sc Aleph} \cite{ref:alephdata}    & & & 22 &  23.0 \\
{\sc Delphi} \cite{ref:delphidata}  &   &  & 51 & 72.2 \\
{\sc Opal} \cite{ref:opaldata,ref:opaleta}  & & & 46 & 192.3 \\
{\sc BaBar} \cite{ref:babardata}     & & & 45  & 44.0 \\ 
{\sc Belle} \cite{ref:belledata}         & & & 78  & 46.8 \\    \hline 
{\sc Hermes} \cite{ref:hermesmult}  & & & 128 & 181.1\\
{\sc Compass} \cite{ref:compassmult} prel.& &       & 398   &  369.9     \\ \hline
{\sc Phenix} \cite{ref:phenixdata}   & & & 15 & 13.9 \\
{\sc Star} \cite{ref:stardata09,ref:stardata13,ref:starcharged06,ref:starratio11}    
                                             &                         &  &  38  & 33.3        \\    
{\sc Alice} \cite{ref:alicedata} \hfill 7~TeV&                        &  &  11 & 32.1        \\ \hline\hline
{\bf TOTAL:} & & & 973 & 1189.5\\
\hline
\end{tabular}
\caption{\label{tab:exppiontab}Data sets used in our NLO global analysis.}
\end{center}
\end{table}

The overall quality of the fit is summarized in Tab.~\ref{tab:exppiontab} 
where we list all data sets included in
our global analysis,
along with their individual $\chi^2$ values.

Firstly, it is worth mentioning that there is a more than twofold increase in the
number of available data points as compared to the original DSS analysis \cite{ref:dss}. 
Secondly, the quality of the global fit has improved dramatically 
from $\chi^2/{\mathrm{d.o.f.}}\simeq 2.2$ for DSS, see Tab.~II in Ref.~\cite{ref:dss}, 
to $\chi^2/{\mathrm{d.o.f.}}\simeq 1.2$ for the current fit. 
A more detailed comparison reveals that the individual $\chi^2$ values for
the SIA data \cite{ref:tpcdata,ref:tassodata,ref:slddata,ref:alephdata,ref:delphidata,ref:opaldata},
which were already included in the DSS fit, have, by and large,
not changed significantly. 
The biggest improvement concerns the SIDIS multiplicities from {\sc Hermes} 
which, in their recently published version \cite{ref:hermesmult}, are now described very well by the 
updated fit. Also, the preliminary charged pion multiplicities from {\sc Compass} \cite{ref:compassmult}
and the new SIA data from {\sc BaBar} \cite{ref:babardata} and {\sc Belle} \cite{ref:belledata}
integrate nicely into the global analysis of parton-to-pion FFs.
Finally, there is some tension among the
$pp$ data sets from RHIC and the LHC, which forced us to introduce a cut $p_T>5\, \mathrm{GeV}$ 
on the pion's transverse momentum in the current fit to accommodate both of them. 
The obtained individual $\chi^2$ values are
all reasonable, as can be inferred from Tab.~\ref{tab:exppiontab}, 
with the new {\sc Alice} data \cite{ref:alicedata} being on the high side, which largely stems from the
penalty for the still sizable normalization shift. This large shift reflects the preference
of the new {\sc Alice} data for a smaller gluon-to-pion FF than extracted by the original DSS fit
based on RHIC {\sc Phenix} data \cite{ref:phenixdata} alone.
As a result of the $p_T$ cut, the number of $pp$ data in the fit for RHIC has decreased as compared to
the DSS analysis. Both the {\sc Brahms} \cite{ref:brahmsdata} and {\sc Star} \cite{ref:starforward} 
results at forward pseudo-rapidities do not pass the $p_T$ cut anymore, and, hence, are excluded from 
the updated fit.

\section{Summary}

We have presented a new, comprehensive global QCD analysis of parton-to-pion fragmentation functions 
at next-to-leading order accuracy including the latest experimental information.
The analyzed data for inclusive pion production in semi-inclusive electron-positron
annihilation, deep-inelastic scattering, and proton-proton collisions span
energy scales ranging from about $1\,\mathrm{GeV}$ up to the mass of the $Z$ boson. 
The achieved, very satisfactory and simultaneous description of all data sets strongly supports the 
validity of the underlying theoretical framework based on pQCD and, in particular, the notion of
factorization and universality for parton-to-pion fragmentation functions.

\section*{Acknowledgments}

The work of RJHP is partially supported by CONACyT, M\'exico. This work was also supported in part by CONICET, ANPCyT, UBACyT, the Research Executive Agency (REA) of the European Union under the Grant 
Agreement number PITN-GA-2010-264564 (LHCPhenoNet)
and the Institutional Strategy of the University of T\"{u}bingen (DFG, ZUK 63).

\section*{References}

\end{document}